# Discovering functional DNA elements using population genomic information: A proof of concept using human mtDNA

Daniel R. Schrider and Andrew D. Kern

Department of Genetics, Rutgers, The State University of New Jersey, Piscataway, NJ 08854

Correspondence: D.R.S. (dan.schrider@rutgers.edu)

## ABSTRACT

Identifying the complete set of functional elements within the human genome would be a windfall for multiple areas of biological research including medicine, molecular biology, and evolution. Complete knowledge of function would aid in the prioritization of loci when searching for the genetic bases of disease or adaptive phenotypes. Because mutations that disrupt function are disfavored by natural selection, purifying selection leaves a detectable signature within functional elements; accordingly this signal has been exploited for over a decade through the use of genomic comparisons of distantly related species. While this is so, the functional complement of the genome changes extensively across time and between lineages, therefore, evidence of the current action of purifying selection in humans is essential. Because the removal of deleterious mutations by natural selection also reduces within-species genetic diversity within functional loci, dense population genetic data have the potential to reveal genomic elements that are currently functional. Here we assess the potential of this approach by examining an ultra-deep sample of human mitochondrial genomes ($n$=16,411). We show that the high density of polymorphism in this dataset precisely delineates regions experiencing purifying selection. Further, we show that the number of segregating alleles at a site is strongly correlated with its divergence across species after accounting for known mutational biases in human mtDNA ($\rho$=0.51; $P$<2.2x10-16). These two measures track one another at a remarkably fine scale across many loci—a correlation that is purely the result of natural selection. Our results demonstrate that genetic variation has the potential to reveal exactly which nucleotides in the genome are currently performing important functions and likely to have deleterious fitness effects when mutated. As more complete human genomes are sequenced, similar power to reveal purifying selection may be achievable in the human nuclear genome.

## INTRODUCTION

Only 1-2% of human genome lies within protein-coding sequence (Lander et al. 2001). Determining the extent to which the remainder of the genome is functional is crucial to our understanding of human biology. A variety of recently developed experimental techniques have aided in the search of non-coding DNA for functional elements (Dunham et al. 2012), however, on their own these techniques can produce a huge number of false positives (Graur et al. 2013). Searches for the evolutionary signature of purifying selection have therefore proved a more fruitful strategy for identifying functional elements; indeed phylogenetic searches comparing sequences of related species have revealed that roughly 5% of the human genome is constrained by natural selection (Chinwalla et al. 2002; Siepel et al. 2005; Lunter et al. 2006; Birney et al. 2007; Davydov et al. 2010), and similar strategies have been used to predict the phenotypic severity of mutations (Stone and Sidow 2005). While whole-genome comparisons aimed at identifying the footprints of selection are highly effective, they have been used primarily to detect elements under constraint for hundreds of millions of years of evolutionary history (Siepel et al. 2005; Davydov et al. 2010). However, the set of functional elements in the genome experiences considerable turnover (Demuth et al. 2006). Comparative genomic techniques will fail in these instances, and particularly for the supremely interesting cases of human-specific gain (Knowles and McLysaght 2009) and loss of function (Wang et al. 2006).

Surveys of genetic diversity within species, on the other hand, have the potential to identify regions currently experiencing purifying selection and that are therefore functional, as purifying selection will remove genetic diversity from such loci. Unfortunately, genetic variation in the human genome is quite sparse, with a comparison of any two homologous chromosomes uncovering <1 single nucleotide polymorphism (SNP) every kilobase (Lander et al. 2001). Sampling more individuals, however, yields additional polymorphisms, and an ultra-deep sample of mitochondrial variation from 16,411 genomes is available on the MITOMAP database (Ruiz-Pesini et al. 2007). These data are extremely polymorphic, with more than one SNP every other base pair on average. This dataset thus serves as an ideal proving ground for the approach of identifying functional constraint using massive amounts of polymorphism data, which will soon be available for nuclear genomes. Here we show that the density of polymorphism in these data closely tracks divergence at a fine scale, implying that these data can indeed be used to reveal the strength of purifying selection in the human mitochondrial genome at a very high resolution. Our results suggest an enormous potential for population genomic data to uncover functional DNA elements, including those not conserved across species.

## RESULTS AND DISCUSSION

We set out to determine the extent to which polymorphism data reveal the strength of purifying selection across the human mitochondrial genome, and downloaded the coordinates of all 8,944 SNPs from MITOMAP (http://www.mitomap.org/MITOMAP). We reasoned that if the density of polymorphism were governed by the amount of purifying selection acting on each site, then SNP density would be correlated with divergence across species, in accordance with expectations under the Neutral model (Kimura 1982). This is indeed what we observe, in the form of a strong correlation between the number of alleles per site and its average negated PhyloP score (Siepel et al. 2006) measuring divergence across vertebrates (Spearman's $\rho$=0.52; $P<2.2\times10^{-16}$). This correlation is also highly significant when averaging polymorphism and divergence within 10 bp adjacent windows ($\rho$=0.50; $P<2.2\times10^{-16}$; Figure 1).

While this observation is consistent with purifying selection both removing diversity and constraining divergence at functional elements, such a pattern could also be generated by variation in the spontaneous mutation rate. It has been shown that mutation rate in the mitochondria varies according to the duration for which a given site remains single stranded on the H strand ($D_{ssH}$) during DNA replication (Reyes et al. 1998). We also find evidence for this in the form a significant correlation between divergence at each site and the duration the site is single-stranded on the H strand during replication, though this correlation is far weaker than that shared between polymorphism and divergence ($\rho$=0.11; $P<2.2\times10^{-16}$). Moreover, after correcting for $D_{ssH}$, the correlation between polymorphism and divergence at individuals sites is essentially unchanged and still highly significant ($\rho$=0.49; $P<2.2\times10^{-16}$). Rather than being driven by a subset of mitochondrial loci, this correlation is significant (at $P<0.05$) in 36/37 genes, and is significant in 35/37 genes after correcting for $D_{ssH}$ (Table 1). This correlation is also far stronger at nonsynonymous than synonymous sites ($\rho$=0.25 for second codon position sites; $P<2.2\times10^{-16}$; $\rho$=0.079 for fourfold degenerate sites; $P=3.6\times10^{-4}$; Figure 2), as expected if purifying selection is a more predominant force at nonsynonymous sites. Finally, the minor allele frequencies of SNPs from mtDB (Ingman and Gyllensten 2006) are correlated with divergence ($\rho$=0.076; $P=4.0\times10^{-6}$), even though variation in mutation rate is not expect to affect allele frequencies. Thus purifying selection uniquely drives patterns of polymorphism in the human mitochondrial genome. This finding supports previous reports that purifying selection is a prominent force in the mitochondrial genome (Rand and Kann 1996; Nielsen and Weinreich 1999; Elson et al. 2004; Stewart et al. 2008), providing further evidence that the hypothesis that drift alone governs mitochondrial evolution (Avise et al. 1987; Palumbi et al. 2001) should be rejected. The patterns we observed are not the result of positive selection, as the fixation of a beneficial mutation through a selective sweep removes all genetic diversity from a non-recombining chromosome (Smith and Haigh 1974). We are thus limited to observing mutations occurring since the most recent sweep.

Having established that patterns of polymorphism across the human mitochondria are largely determined by purifying selection, we sought to determine the resolution at which these data reveal the strength of selection acting on particular sites in the genome. We examined patterns of SNP diversity and divergence in 5 bp sliding windows across each gene in the mitochondrial genome, observing the extent to which the two measures mirror one another on a small scale. In particular, within each window we calculated the average SNP density per base pair and the average probability that the site is not conserved across vertebrates according to phastCons (Siepel et al. 2005). We find that for many loci, tRNA genes in particular, these two measures track one another to a surprising extent (e.g. the phenylalanine and tryptophan tRNA genes shown in Figure 3; the remaining tRNA genes are shown in supplementary figure S1). This result demonstrates that SNP density has the ability to reveal the strength of selection at a surprisingly detailed resolution—on the scale of a few base pairs. We therefore sought to use SNP density to predict function at a fine scale via a hidden Markov model (HMM; Rabiner 1989). Using a similar strategy as phastCons (Siepel et al. 2005), we learned a two-state HMM (constrained vs. unconstrained) where the observation for each site in the genome is the number of alleles at the site. We then used this HMM to predict constrained regions to which we refer as mitoPopCons elements. There is extremely strong overlap between mitoPopCons elements obtained from polymorphism data and phastCons elements predicted from divergence (*P*-value<0.0001; Figure 4; Methods). SNP diversity therefore has the ability to accurately predict function at a fine scale in the human mitochondria.

We have shown that ultra-dense polymorphism data can be used to accurately detect functional nucleotides in the human mitochondrial genome, potentially at the level of the individual base pair, while sidestepping limitations of phylogenetic approaches. This result suggests that as whole-genome sequencing becomes more ubiquitous, it will become possible to perform such high-resolution prediction in the nuclear genome as well. Approaches such as ours will then have an enormous impact on biological research, allowing for the discovery of the complete set of functional elements in the human genome and the degree to which new mutations at each site are deleterious. Such efforts will vastly improve predictions of the phenotypic impact of mutations occurring in humans and will prioritize searches for disease-causing mutations. This information will also reveal species-specific changes in selective pressures at the resolution of individual nucleotides, greatly improving our understanding of how the functional components of genomes evolve.

## METHODS

We converted all coordinates from obtained from MITOMAP and mtDB from the CRS to the human reference genome's (GRCh37) coordinate space by aligning the two mitochondrial sequences using MUSCLE (Edgar 2004). We downloaded phyloP scores for each mitochondrial site from the UCSC Genome Browser Database (Meyer et al. 2013). Because there are no allele frequency data available for MITOMAP SNPs, we used the number of alleles at each site as our measure of diversity. For the correlation of minor allele frequency with divergence, we used frequencies of biallelic SNPs from mtDB. We measured the duration of single-strandedness ($D_{ssH}$) following Chong and Mueller (Chong and Mueller 2013), using the coordinates of the light-strand origin of replication from MITOMAP. For the $D_{ssH}$ analyses we omitted sites within the control region. To control for the $D_{ssH}$-divergence correlation, we fit a generalized linear model treating the number of segregating alleles at each site as a Poisson distributed response variable and with $D_{ssH}$ value as the predictive variable using R. We then calculated the correlation between PhyloP scores and the polymorphism residuals. The HMM analysis was performed in MATLAB with the same transition matrix as used to learn the phastCons HMM for the UCSC Genome Browser (Kent et al. 2002). We then used Baum-Welch training to learn the emission parameters only for the constrained and unconstrained states by using large number of pseudocounts for transition matrix to ensure it remained invariant during training. Next, we predicted constrained elements by using the Viterbi algorithm. We tested for significant overlap between our HMM predictions and mammalian phastCons elements from the UCSU Genome Browser by counting the number of base pairs constrained according to both methods, and compared this count to those obtained after permuting the coordinates of our predictions. This was repeated for 10,000 permutations to obtain a *P*-value.

## ACKNOWLEDGEMENTS

D. R. S. was supported by the National Institutes of Health under Ruth L. Kirschstein National Research Service Award F32 GM105231-01, and A. D. K. was supported in part by Rutgers University and National Science Foundation Award MCB-1161367.

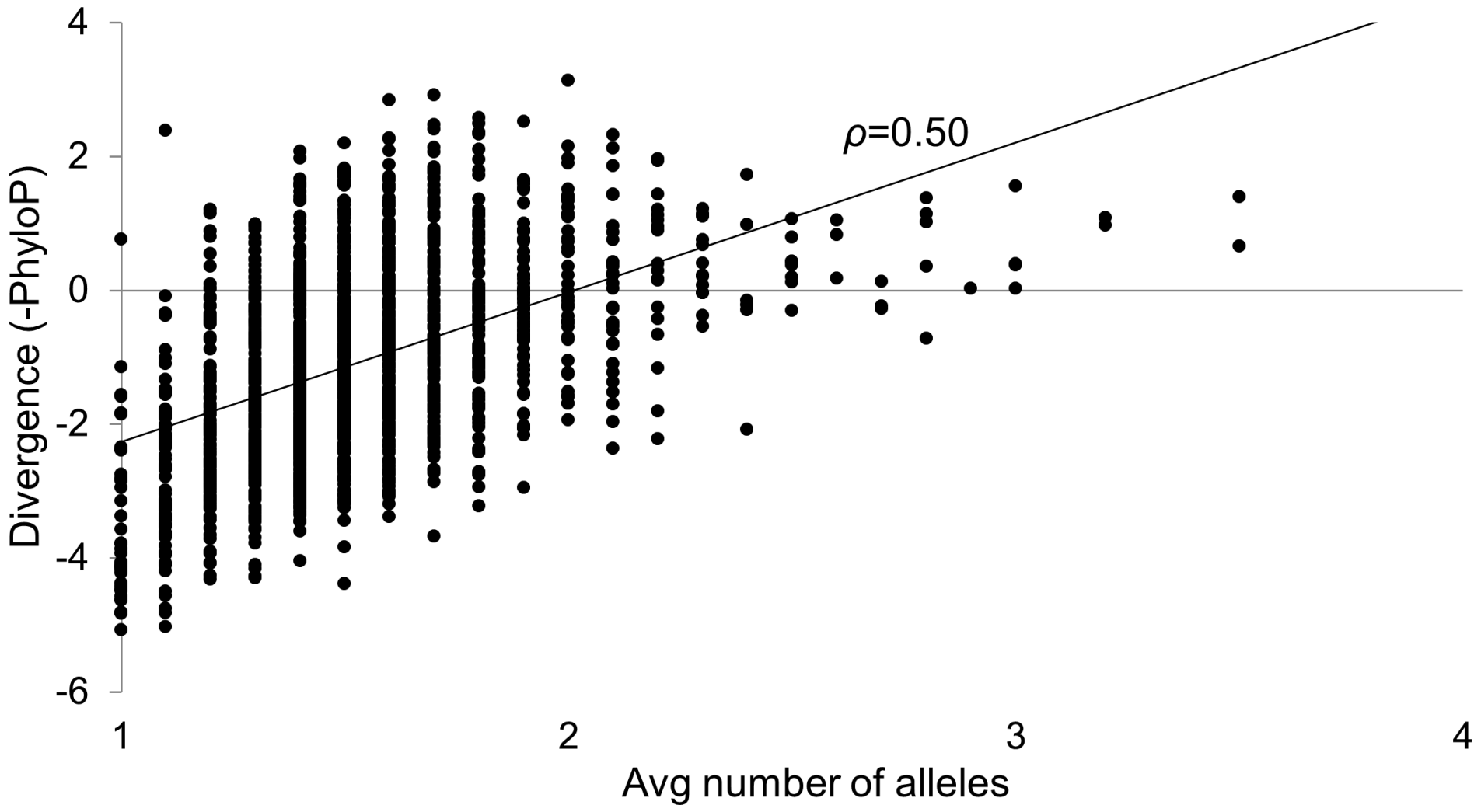

**Figure 1: The correlation between polymorphism and divergence in the human mitochondrial genome.** The average number of alleles per base pair in 10 bp windows is shown on the x-axis, and divergence as measured by the negated phyloP score is shown on the y-axis.

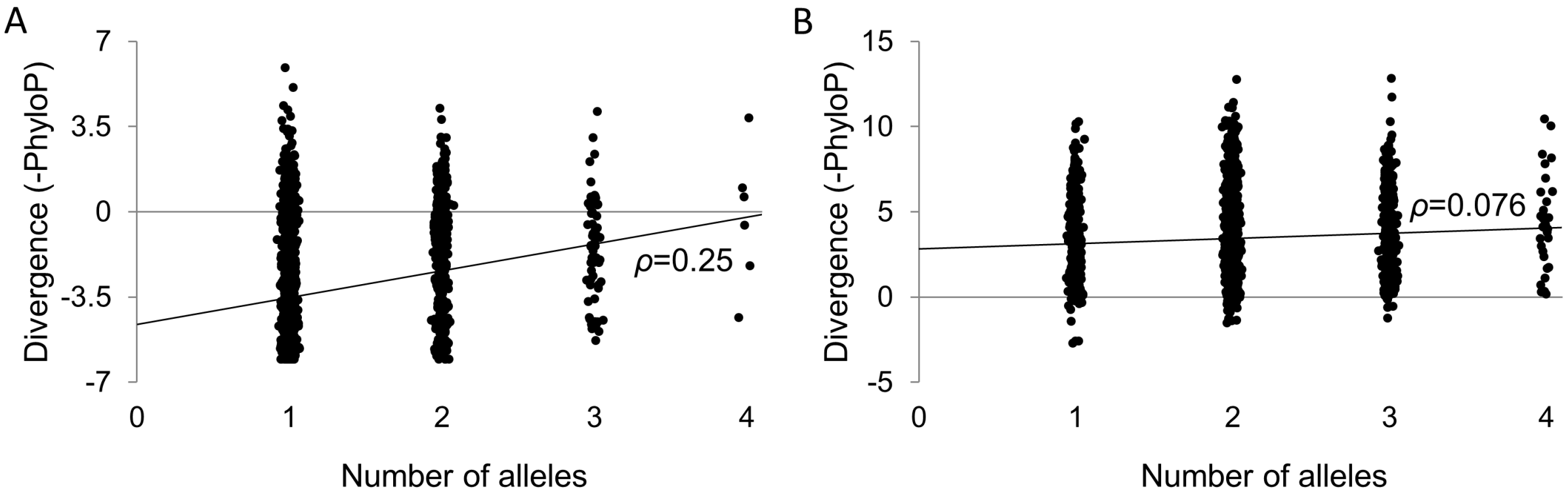

**Figure 2: The correlation between polymorphism and divergence at nonsynonymous and synonymous sites.** The number of alleles observed at each site is shown on the x-axis and divergence (negative phyloP score) is shown on the y-axis at (A) 2nd codon position (nonsynonymous) sites and (B) 4-fold degenerate (synonymous) sites. We added noise to the number of alleles in order to reveal the density of sites along the y-axis.

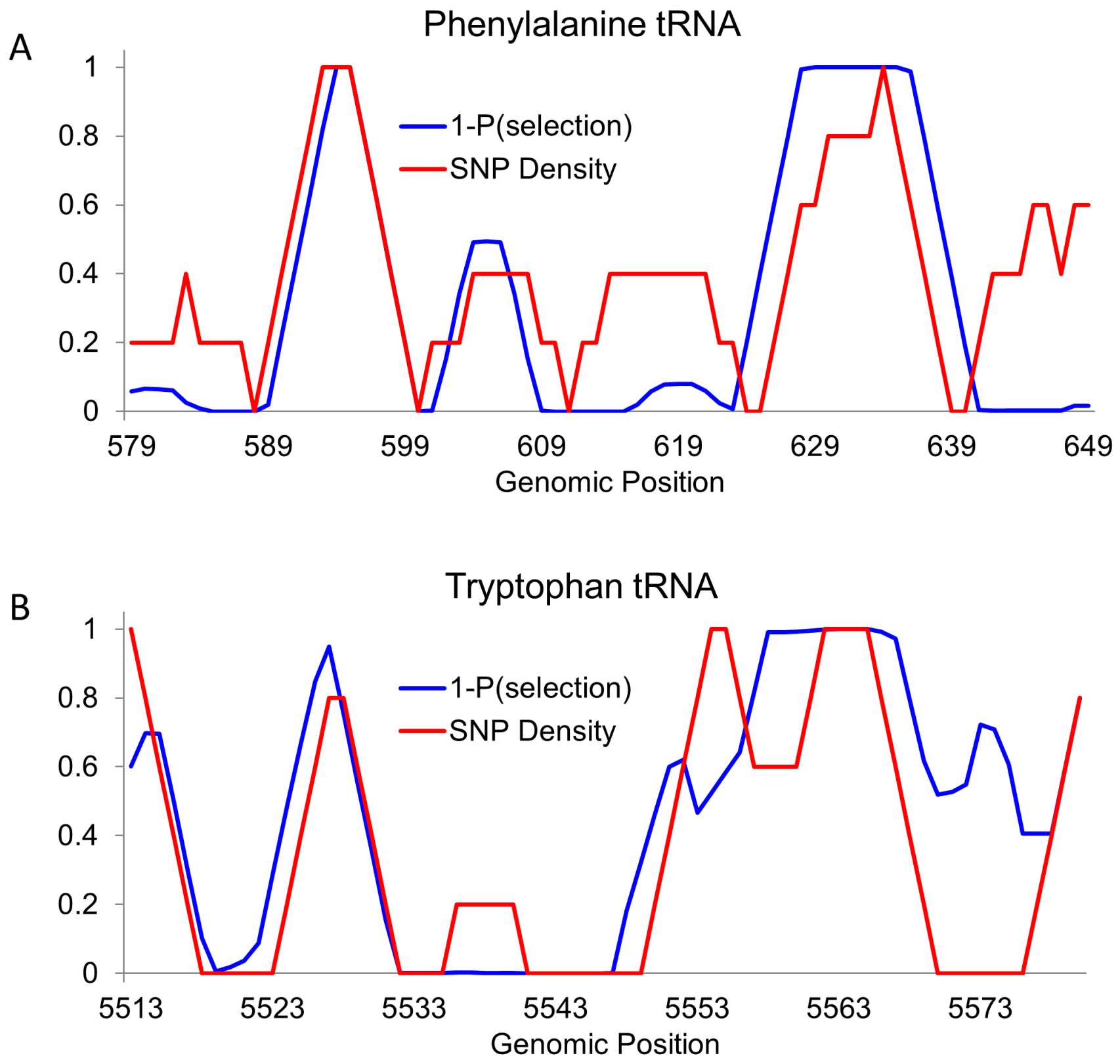

**Figure 3: The probability of polymorphism versus the probability of unconstrained evolution across vertebrates.** (A) 5bp sliding genomic windows showing SNP density (blue) and one minus the probability of conservation across vertebrates (red) according to PhastCons(Siepel et al. 2005) across the Phenylalanine tRNA gene. (B) The same plot for the Tryptophan tRNA gene.

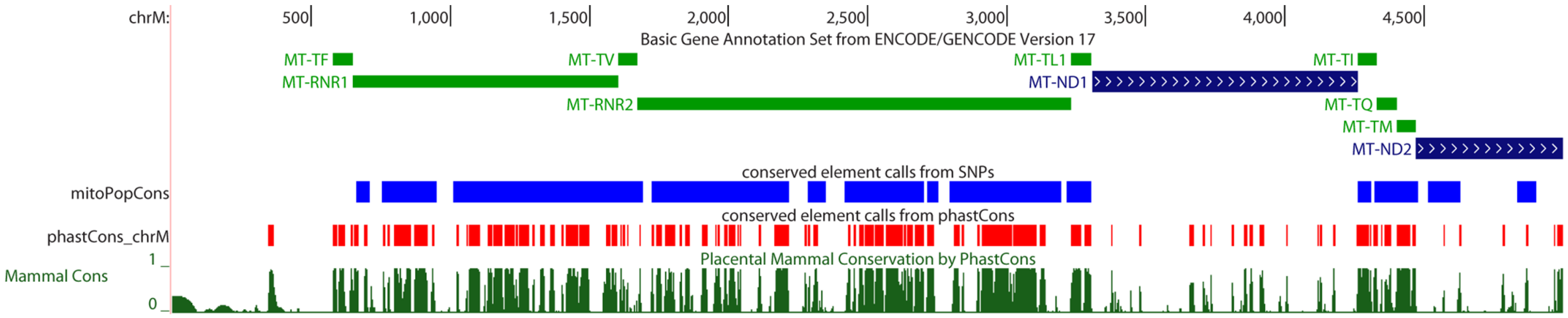

**Figure 4: Comparison of conserved elements called from phylogenetic data and those called from population genetic data.** This image form the UCSC Genome Browser shows positions 1-5,000 of the human mitochondrial genome. Conserved elements called from the polymorphism-based HMM (mitoPopCons) appear in blue, while phastCons elements obtained from a comparison of mammalian genomes appear in red. PhastCons conservation probabilities are shown at the bottom in green. Gene locations are shown at the top.

**Table 1: Gene-specific correlations between SNP density and negative phyloP score.**

| Gene Name | Gene Description | Gene Start (hg19) | Gene End (hg19) | Spear man's $\rho$ | *P*-value | Spearman's $\rho$ (correcting for $D_{SSH}$) | *P*-value (correcting for $D_{SSH}$) |
|---|---|---|---|---|---|---|---|
| MT-TF | tRNA phenylalanine | 579 | 649 | -0.484 | 1.90E-05 | -0.468948 | 3.71E-05 |
| MT-RNR1 | 12S ribosomal RNA | 650 | 1603 | -0.429 | <2.20E-16 | -0.4169046 | <2.20E-16 |
| MT-TV | tRNA valine | 1604 | 1672 | -0.336 | 0.004783 | -0.326022 | 0.006261 |
| MT-RNR2 | 16S ribosomal RNA | 1673 | 3230 | -0.443 | <2.20E-16 | -0.4435798 | <2.20E-16 |
| MT-TL1 | tRNA leucine 1 | 3231 | 3305 | -0.301 | 0.008593 | -0.29837 | 0.00932 |
| MT-ND1 | NADH Dehydrogenase subunit 1 | 3308 | 4263 | -0.542 | <2.20E-16 | -0.5149936 | <2.20E-16 |
| MT-TI | tRNA isoleucine | 4264 | 4332 | -0.269 | 0.02551 | -0.2523745 | 0.03643 |
| MT-TQ | tRNA glutamine | 4330 | 4401 | -0.463 | 4.19E-05 | -0.476385 | 2.34E-05 |
| MT-TM | tRNA methionine | 4403 | 4470 | -0.291 | 0.01611 | -0.2880727 | 0.01721 |
| MT-ND2 | NADH dehydrogenase subunit 2 | 4471 | 5512 | -0.516 | <2.20E-16 | -0.461188 | <2.20E-16 |
| MT-TW | tRNA tryptophan | 5513 | 5580 | -0.509 | 9.42E-06 | -0.4626812 | 7.11E-05 |
| MT-TA | tRNA alanine | 5588 | 5656 | -0.429 | 0.0002381 | -0.4412621 | 0.0001475 |
| MT-TN | tRNA asparagine | 5658 | 5730 | -0.378 | 0.0009899 | -0.3765626 | 0.001025 |
| MT-TC | tRNA cysteine | 5762 | 5827 | -0.596 | 1.26E-07 | -0.5709837 | 5.55E-07 |
| MT-TY | tRNA tyrosine | 5827 | 5892 | -0.352 | 0.00371 | -0.3486535 | 0.004118 |
| MT-CO1 | Cytochrome c oxidase subunit I | 5905 | 7446 | -0.633 | <2.20E-16 | -0.6186776 | <2.20E-16 |
| MT-TS1 | tRNA serine 1 | 7447 | 7515 | -0.627 | 8.31E-09 | -0.6107229 | 2.51E-08 |
| MT-TD | tRNA aspartic acid | 7519 | 7586 | -0.365 | 0.002186 | -0.320253 | 0.007758 |
| MT-CO2 | Cytochrome c oxidase subunit II | 7587 | 8270 | -0.573 | <2.20E-16 | -0.5291891 | <2.20E-16 |
| MT-TK | tRNA lysine | 8296 | 8365 | -0.331 | 0.005158 | -0.2809919 | 0.01846 |
| MT-ATP8 | ATP synthase F0 subunit 8 | 8367 | 8573 | -0.266 | 0.0001042 | -0.2637231 | 0.0001233 |
| MT-ATP6 | ATP synthase F0 subunit 6 | 8528 | 9208 | -0.389 | <2.20E-16 | -0.3823366 | <2.20E-16 |
| MT-CO3 | Cytochrome c oxidase subunit III | 9208 | 9991 | -0.538 | <2.20E-16 | -0.5314281 | <2.20E-16 |
| MT-TG | tRNA glycine | 9992 | 10059 | -0.396 | 0.0008191 | -0.3404812 | 0.004497 |
| MT-ND3 | NADH dehydrogenase subunit 3 | 10060 | 10405 | -0.510 | <2.20E-16 | -0.5000739 | <2.20E-16 |
| MT-TR | tRNA arginine | 10406 | 10470 | -0.470 | 7.66E-05 | -0.4415206 | 0.0002317 |
| MT-ND4L | NADH dehydrogenase subunit 4L | 10471 | 10767 | -0.491 | <2.20E-16 | -0.5133884 | <2.20E-16 |
| MT-ND4 | NADH dehydrogenase subunit 4 | 10761 | 12138 | -0.561 | <2.20E-16 | -0.5419698 | <2.20E-16 |
| MT-TH | tRNA histidine | 12139 | 12207 | -0.277 | 0.02112 | -0.1786413 | 0.1419 |
| MT-TS2 | tRNA serine2 | 12208 | 12266 | -0.470 | 0.0001712 | -0.489146 | 8.45E-05 |
| MT-TL2 | tRNA leucine2 | 12267 | 12337 | -0.374 | 0.001311 | -0.3543069 | 0.002434 |
| MT-ND5 | NADH dehydrogenase subunit 5 | 12338 | 14149 | -0.532 | <2.20E-16 | -0.5141183 | <2.20E-16 |
| MT-ND6 | NADH dehydrogenase subunit 6 | 14150 | 14674 | -0.500 | <2.20E-16 | -0.4776221 | <2.20E-16 |
| MT-TE | tRNA glutamic acid | 14675 | 14743 | -0.347 | 0.003468 | -0.3476731 | 0.003421 |
| MT-CYB | Cytochrome b | 14748 | 15888 | -0.551 | <2.20E-16 | -0.5025513 | <2.20E-16 |
| MT-TT | tRNA threonine | 15889 | 15954 | -0.560 | 1.00E-06 | -0.4354723 | 0.0002578 |
| MT-TP | tRNA proline | 15957 | 16024 | -0.103 | 0.4044 | -0.1870294 | 0.1267 |

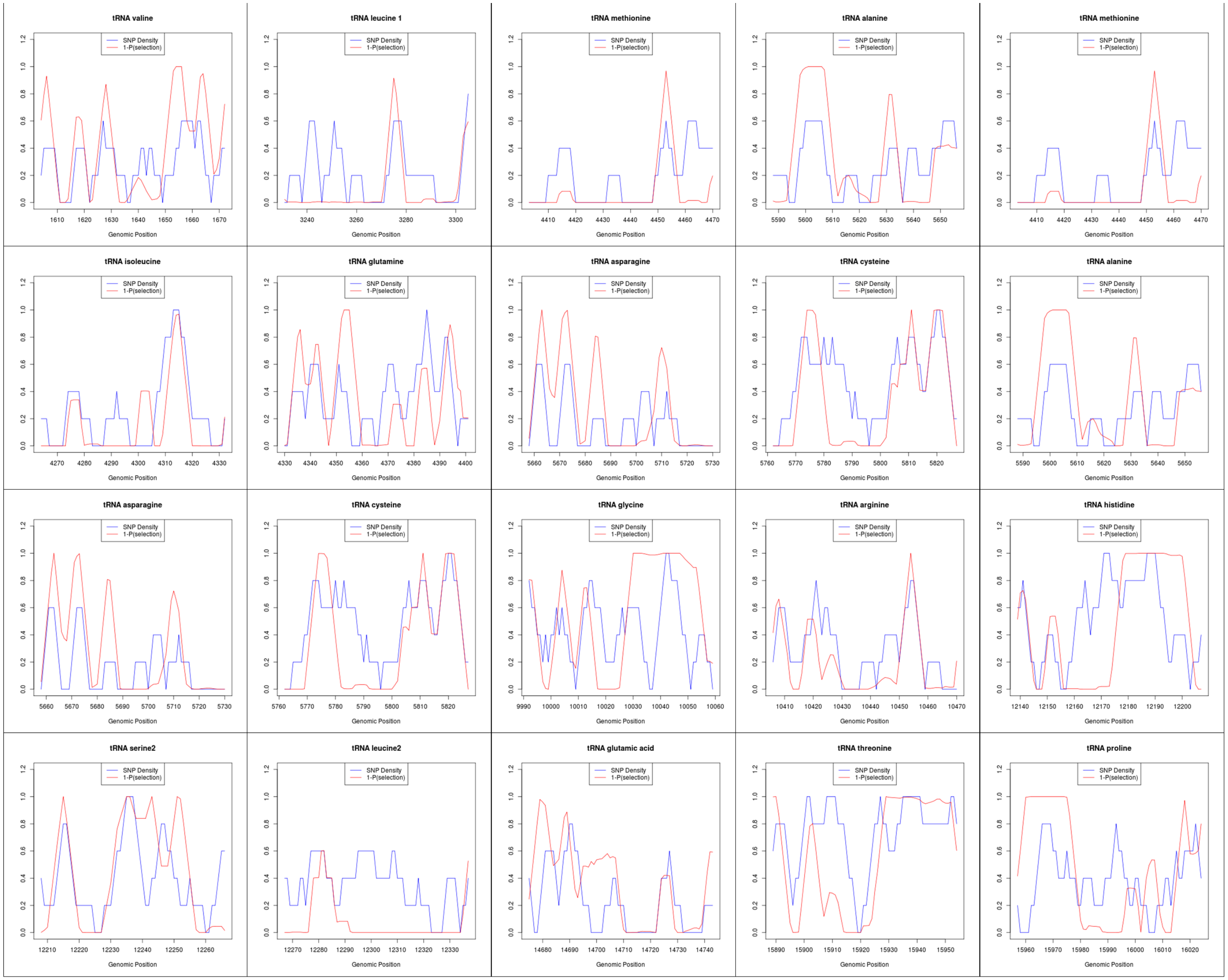

**Supplementary figure S1: Polymorphism versus divergence in 5bp adjacent windows across the 20 remaining mitochondrial tRNA genes.** Phenylalanine and Tryptophan are shown in Figure 3.